# Emerging Cloud Computing Security Threats

**Kamal A. Ahmat**

Department of Information Technology

City University of New York

kamal.ahmat.us@ieee.org

**Abstract**

Cloud computing is one of the latest emerging innovations of the modern internet and technological landscape. With everyone from the White house to major online technological leaders like Amazon and Google using or offering cloud computing services it is truly presents itself as an exciting and innovative method to store and use data on the internet.

In this paper, we discuss some of the most key evolving threats on Cloud systems and related services. Then, we provide some analysis results to quantify the impact of these threats.

## 1 INTERODUCTION

Cloud computing is one of the latest emerging innovations of the modern internet and technological landscape. with everyone from the White house to major online technological leaders like Amazon and Google using or offering cloud computing services it is truly presents itself as an exciting and innovative method to store and use data on the internet. By offering software, storage and other services via an online account, Cloud providers can greatly reduce costs for small and large businesses or startups by giving them access to features that may be very costly otherwise.

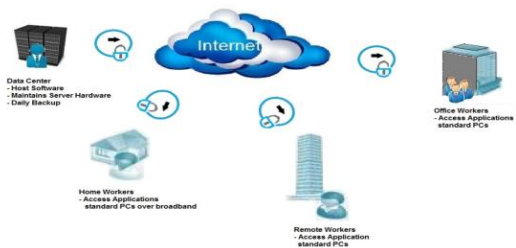

*Fig 1 Cloud computing diagram*

By offering software, applications, storage and other services online account, Cloud providers can greatly reduce costs for small/large businesses and startups by giving them access to advanced features that may be very costly otherwise and far beyond their means to obtain or maintain. They can simplify a company's it infrastructure by providing turnkey solutions to non technical staff such as CRMS (Customer relationship management systems), leads, telephony, accounting apps, contacts and databases. Cloud services like Amazon Cloud and Sales Genie currently do provide online CRM, sales leads, database and payment services all in one affordable account depending on the customers' needs.

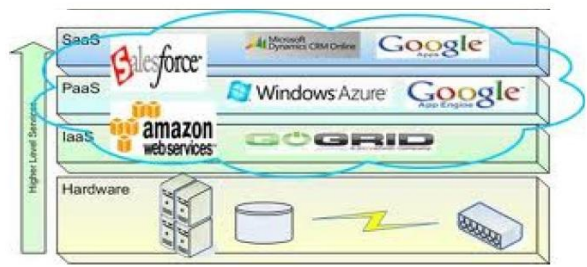

*Fig 2 Diagram of architecture based on Cloud Services*

## 2 RELATED WORK

Agrawal *et al.*[1] analyzed the design choices that allowed modern scalable data management systems to achieve orders of magnitude higher levels of scalability compared to traditional databases. J. Morin *et al.* [2] identified Cloud Computing Security issues and their corresponding challenges, proposing to use risk and Service Level Agreement (SLA) management as the basis for a service level framework to improve governance, risk and compliance in cloud computing environments.

N. Mayer *et al.* [3] defined the set of concepts and relationships taking a place in the ISSRM domain within a UML class diagram. The outcome of their work is the enrichment of the class diagram with attributes representing the elicited metrics.

S. Carlin *et al.* [4] proved that one of the biggest security worries with the cloud computing model is the sharing of resources (multitenancy). So, new security techniques need to be developed and older security techniques needed to be radically tweaked to be able to work with the clouds architecture. Plugging in existing security technology will not work because this new delivery model introduces new changes to the way in which we access and use computer resources.

Wang *et al.* [5] presented a brief summary on the analysis of current gaps and new trends in cloud computing research based on extant information systems literature, industry reports, and practical experience reflections. Additionally, it highlights the significance of cloud computing and its implications for practitioner and academics.

Zissis and Lekkas [6] concluded that; firstly to evaluate cloud security by identifying unique security requirements and secondly to attempt to present a viable solution that eliminates these potential threats. This paper proposes introducing a Trusted Third Party, tasked with assuring specific security characteristics within a cloud environment. The proposed solution calls upon cryptography, specifically Public Key Infrastructure operating in concert with SSO and LDAP, to ensure the authentication, integrity and confidentiality of involved data and communications.

Kuyoro *et al.* [7] introduced a detailed analysis of the cloud computing security issues and challenges focusing on the cloud computing types and the service delivery types. In this paper key security considerations and challenges which are currently faced in the Cloud computing are highlighted. Cloud computing has the potential to become a frontrunner in promoting a secure, virtual and economically viable IT solution in the future.

Mishra *et al.* [8] brought their own security concerns to the already large list of cloud computing. As multi-tenancy, virtualization comes with its own issues. The hypervisor provides a new attack surface to be compromised; and the virtual network enables a malicious VM to perform attacks on other VMs avoiding traditional network security controls. The movement to the Cloud could mean an improvement in security to many organizations. New robust security controls will be required in order to assure proper security with the de-parameterization, and to be compliant with the everyday more strict laws and regulations.

Finally, [9] presented a framework for web-based interactive scalable network visualization. WiNV enables a new class of rich and scalable interactive cross-platform capabilities for visualizing large-scale networks natively in a user's browser. This allows visualizing cloud security configurations and detecting potential holes.

## 3 SECURITY CHALLENGES

In spite of all the many advantages, due issues such as lack of standardization, there are unfortunately some drawbacks and security issues that can arise from using cloud computing and SaaS services (Software-as-a-service). This paper describes some of the major security risks facing cloud providers and mentions several recognized by the Cloud Security Alliance (CSA). We focus on the following threats:

1. Cyber attacks and hacking of sensitive information.
2. Illegal local network access from cloud services.
3. Stolen information from cloud computing employees.
4. Attacks from other customers.
5. Adherence and compliance of providers to security standards.
6. Data loss.
7. Data Segregation from other customers.
8. Security culture among providers.
9. Evolving threats that may target clouds.
10. Privacy concerns.

SaaS undeniably provides savings and other advantages for end users but we now endeavor to address its risks and possible solutions to make it more secure for customers.

### 3.1 Cyber attacks and hacking of sensitive information

Cloud computing is clearly an advancement of other web services like web hosting and online storage and so it faces many of the same risks and from hackers and cyber thieves. Hackers who are able to break into a public or private cloud computing environment can steal sensitive information from many different users and either use or sell that information. Information such a credit card numbers, financial records, software and

(2)

reports are obviously stored online by many users and are constantly at risk of being stolen. CSP's (Cloud service providers) therefore are obligated to continually stay vigilant to keep this information safe, consequently, user data safety depends on a CSP's individual safety level and culture.

### 3.2 Illegal local network access from cloud services

If hackers are successful in infiltrating a cloud they can further cause damage by planting viruses and worms which could further seek to access a user's office or personal network. Most CSP's fortunately do provide antiviruses and firewalls' but with the evolving nature of online threats; a compromised cloud could be a near disaster for end users and providers both.

### 3.3 Stolen information from cloud computing employees

While CSP's may be successful in out threats external to the cloud, it is a bit harder to keep out internal threats such as a bad employee intent in copying sensitive data to a flash drive or other storage device. It firms continually face this problem and naturally obvious solutions would include limiting reading and copying permissions within the CSP and internal monitoring if the cloud. Limiting physical access to sensitive data and proper background checks is also vital.

### 3.4 Attacks from other customers

It is quite possible that some users of Cloud services sign up simply to steal information from other users. CSP's therefore must monitor the usage of its customers so that illegal usage of the cloud by users does not lead to direct or accidental leakage of data. Proper screening and validation of potential users can help to reduce this risk.

### 3.5 Adherence and compliance of providers to security standards

The Cloud Security Alliance (CSA) helps Cloud providers by providing key tips to prevent the recurring threats that all CSP's face, many of which are mentioned as we discuss each threat in this paper. other bodies such as the open Cloud Consortium are endeavoring to set guidelines and policies that can help cloud providers to grow and meet all security threats.

### 3.6 Data loss

Some malware and viruses are solely interested in destruction of data which can be fatal in the cloud. Whether Data loss is by hardware issues, malware or other causes, this particular threat can reduced by proper backup and disaster recovery by Cloud providers.

### 3.7 Data Segregation from other customers

Improper partitioning of users and their data can also cause vulnerability and other problems in the cloud.

### 3.8 Security culture among providers

In a survey conducted by the Phenom institute, many Cloud providers that participated felt that the customer was more responsible for security in the cloud than the provider. other findings include that Cloud Security Providers feel the main advantages they offer users are reduced costs and savings, that few of them (26%of participants in the USA) have full time cloud security personnel and many Cloud providers (over 65%) do not encrypt their customers data. while many of the participants do have features like firewalls and antivirus security risks like unauthorized customer logins are not as closely monitored by some Cloud providers. A few Cloud Security providers do claim to offer the highest level of security for their customers with about 10% stating that they offer Security as a Service in the Cloud. Security as a Service is the concept of outsourcing security needs to an outside party and is likely to see much growth in the cloud especially for users who feel unable to safeguard their networks to the highest level. Figure 3 shows the percentages of responsibility for ensuring the security of Cloud resources by Cloud providers.

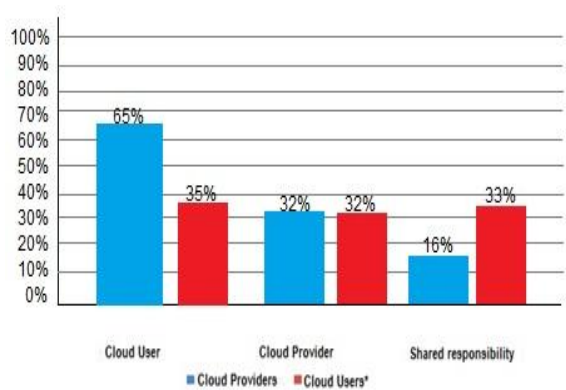

*Fig 3 Data from cloud user study*

## 4 EVOLVING THREATS THAT MANY TARGET CLOUD

Cloud services are fairly new and growing evolving field therefore new and evolving threats, malware and attacks will no doubt emerge that target specific clouds and their vulnerabilities So even though a cloud is fairly secure today, malware and other threats may emerge in the future that are able to compromise it- as with any online service. Cloud providers who do not anticipate and stay aware of new threats could easily fall prey to new security risks as they emerge.

## 5 PRIVACY CONCERNS

Finally one of the drawbacks of the Cloud is that customers may have to trust their providers with sensitive, risky and potentially damaging information from a legal standpoint. While legal constraints would require Cloud providers to keep all user data in trust, private data could leak or be required by authorities in the case of legal proceedings.

## 6 CONCLUSION

This paper addressed security risks facing cloud providers and mentioned several recognized by the Cloud Security Alliance (CSA). They include:

- Cyber attacks and hacking of sensitive information
- Illegal local network access from cloud services.
- Stolen information from cloud computing employees.
- Attacks from other customers
- Adherence and compliance of providers to security standards,
- Data loss
- Data Segregation from other customers.
- Security culture among providers.
- Evolving threats that may target clouds.
- Privacy concerns